\documentclass[twocolumn,showpacs,nofootinbib,preprintnumbers,amsmath,amssymb,showkeys,superscriptaddress]{revtex4-1}
\usepackage{epsfig}
\usepackage{dsfont}
\usepackage{color}
\usepackage{graphicx}
\usepackage{dcolumn}
\usepackage{bm}
\usepackage{physics}

\newcommand \beq{\begin{equation}}
\newcommand \eeq{\end{equation}}
\newcommand \bea{\begin{align}}
\newcommand \eea{\end{align}}

\newcommand\eqn[1]{(\ref{#1})}      
\newcommand\Eqn[1]{Eq.~(\ref{#1})}  
\newcommand\Fig[1]{Fig.~\ref{#1}}  

\newcommand{\cD}{{\cal D}}

\begin{document}
\allowdisplaybreaks

\title{Stability of de Sitter spacetime against infrared quantum scalar field fluctuations}

\author{G. Moreau} 
\author{J. Serreau}
\affiliation{APC, AstroParticule et Cosmologie, Universit\'e Paris Diderot, CNRS/IN2P3, CEA/Irfu, Observatoire de Paris, Sorbonne Paris Cit\'e, 10, rue Alice Domon et L\'eonie Duquet, 75205 Paris Cedex 13, France.}
\date{\today}

\begin{abstract}
We study the backreaction of superhorizon fluctuations of a light quantum scalar field on a classical de Sitter geometry by means of the Wilsonian renormalisation group. This allows us to treat the gravitationally amplified fluctuations in a nonperturbative manner and to analytically follow the induced renormalisation flow of the spacetime curvature as long wavelength modes are progressively integrated out. Unbounded loop corrections in the deep infrared are eventually screened by nonperturbative effects which stabilise the geometry. 
\end{abstract}

\maketitle


Despite more than half a century of efforts (and numerous progresses), the extreme smallness of the measured cosmological constant in Planck units remains a major open issue in Physics, which directly questions our fundamental understanding of gravity \cite{Weinberg:1988cp,Padmanabhan:2002ji,Martin:2012bt}. In the language of quantum field theory, the problem appears as that of an unnatural fine tuning between the---{\it a priori} arbitrary---bare cosmological constant, allowed by all symmetries known to date, and the---inevitably large---quantum contribution from vacuum fluctuations of the Standard Model fields. Some potential key players are, however, missing in this picture. Obvious ones are the quantum fluctuations of the gravitational field itself, which have been suggested early on as a possible solution of the puzzle \cite{Polyakov:1982ug,Antoniadis:1991fa,Tsamis:1992sx}. There coud also be important backreaction effects of the quantum fluctuations in the matter (nongravitational) sector \cite{Myhrvold:1983hx,Ford:1984hs,Mottola:1984ar,Antoniadis:1985pj,Mazur:1986et}.   

Fortunately, a complete theory of quantum gravity might not be needed to address this question, which concerns the far infrared sector of the theory \cite{Tsamis:1992sx}. A semiclassical description, with quantum fluctuations self-consistently coupled to a (dynamical) classical gravitational field through Einstein's equations, already provides a framework for a possible solution. In the absence of quantum fluctuations, the classical solution to Einstein's equations with a positive (negative) cosmological constant is the maximally symmetric de Sitter (anti-de Sitter) geometry. In this context, it has been long  suggested that the classical de Sitter geometry may be unstable against quantum fluctuations \cite{Polyakov:1982ug,Ford:1984hs,Mottola:1984ar,Antoniadis:1985pj,Mazur:1986et,Antoniadis:1991fa,Tsamis:1992sx}, a possibility which has received a renewed interest in the past two decades \cite{Onemli:2002hr,Antoniadis:2006wq,Polyakov:2007mm,Krotov:2010ma,Marolf:2010zp,Hollands:2010pr,Boyanovsky:2011xn,Tsamis:2011ep,Anderson:2013ila,Akhmedov:2012dn,Akhmedov:2013xka,Kaya:2013bga,Akhmedov:2017ooy,Markkanen:2017abw,Anderson:2017hts,Miao:2017vly}. In standard cosmological coordinates, the qualitative picture goes as follows. The accelerated spacetime expansion pulls appart particle-antiparticle pairs from the quantum vacuum. The self-gravitation of the latter may then slows the expansion down, resulting in an effective decay of the spacetime curvature, that is, of the effective cosmological constant.

The route to establish this scenario is, however, paved with serious technical difficulties. Leave alone the hard task of properly computing graviton loop corrections in a de Sitter geometry \cite{Tsamis:1996qk,Senatore:2009cf,Giddings:2010nc}, even the case of a simple scalar quantum field is far from trivial. First, convincingly assessing the question of a possible instability (beyond a linear analysis) requires one to actually control loop corrections of the scalar field in a non de Sitter geometry, with less symmetries, where calculations are technically more involved \cite{Koivisto:2010pj,Markkanen:2013nwa,Glavan:2017jye}. Second, the strong particle production from the de Sitter gravitational field actually results in dramatically amplified quantum fluctuations and, in turn, in infrared divergent loop contributions \cite{Tsamis:2005hd,Weinberg:2005qc}. The latter signal a breakdown of perturbation theory and a proper treatment of quantum contributions thus requires resummation techniques or genuine nonperturbative approaches \cite{Starobinsky:1994bd,Burgess:2009bs}.

Here, we propose a novel perspective on the problem of backreaction based on the recent developments of nonperturbative renormalisation group (NPRG) techniques in de Sitter spacetime \cite{Kaya:2013bga,Serreau:2013eoa,Guilleux:2015pma,Gonzalez:2016jrn,Prokopec:2017vxx}. This consists in progressively integrating out the infrared sector of the theory and it has been been shown to efficiently capture the nonperturbative infrared dynamics of quantum scalar fields in this context. We adapt the method to the semiclassical problem at hand by including a classical gravitational field, self-consistently determined through the semiclassical Einstein's equations at each renormalisation group (RG) scale. We can, thus, follow the RG flow of the effective spacetime curvature as infrared, superhorizon fluctuations are progressively integrated out.

This approach offers various technical advantages. First, of course, we fully capture the nonperturbative character of the problem. Second, we can consistently work in de Sitter spacetime since we follow the RG trajectory of the theory, instead of its time evolution. Third, we can focus specifically on the role of the infrared modes, the ultraviolet contributions being absorbed in the initial conditions of the RG flow. Finally, as we shall see below, most calculations can be performed analytically thanks to the simple nature of the RG flow in the far infrared.

We consider the theory of a quantum scalar field $\hat\varphi$ coupled to a classical gravitational field $g_{\mu\nu}$, described by the generating functional
\beq \label{eq:measure}
 e^{-iW_\kappa[J,\,g]}=\int\cD\hat\varphi e^{iS[\hat\varphi,g]+i\Delta S_\kappa[\hat\varphi,g]-iJ\cdot\hat \varphi},
\eeq
where $S$ is the classical action, $J\cdot\hat\varphi\equiv\int_xJ(x)\hat\varphi(x)$, and where the quadratic modification of the action
\beq
 \Delta S_\kappa[\varphi,g]=\frac{1}{2}\int_{x,y}\!\!R_\kappa(x,y)\varphi(x)\varphi(y)
\eeq 
plays the role of an infrared cutoff, which suppresses fluctuations of wavelength larger than $1/\kappa$ (in the sense of the  metric $g_{\mu\nu}$) from the path integral.
Here, $\int_x=\int d^4 x\sqrt{-g}$  is the invariant measure and we define the corresponding functional trace as ${\rm Tr}_g{\cal F}=\int_x{\cal F}(x,x)$. 
The regularised effective action $\Gamma_\kappa[\varphi,g]$, with $\varphi\equiv\ev{\hat\varphi}$, is defined through the modified Legendre transform
$ \Gamma_\kappa[\varphi,g]+\Delta S_\kappa[\varphi,g]+W_\kappa[J,g]=J\cdot\varphi$,
and interpolates between the microscopic action for $\kappa$ large compared to any other scale in the problem, $\Gamma_{\kappa\to\infty}=S$, and the usual effective action $\Gamma_{\kappa\to0}=\Gamma$. It satisfies the exact flow equation \cite{Wetterich:1992yh,Delamotte:2007pf} 
\beq\label{eq:Wett}
 \partial_\kappa\Gamma_\kappa[\varphi,g]=\frac{i}{2}{\rm Tr}_g\!\left\{\partial_\kappa R_\kappa[g]\cdot(\Gamma_\kappa^{(2)}[\varphi,g]+R_\kappa[g])^{-1}\right\},
\eeq
where $\Gamma_\kappa^{(2)}(x,y)=[g(x)g(y)]^{-{1\over2}}\delta^2\Gamma_\kappa[\varphi,g]/\delta\varphi(x)\delta\varphi(y)$. The physical point in field/metric space is given, at each RG scale $\kappa$, by the extremisation conditions
\beq\label{eq:geom}
 \left.\frac{\delta \Gamma_\kappa[\varphi,g]}{\delta \varphi(x)}\right|_{\varphi_\kappa,g_\kappa}=0\,,\quad \left.\frac{\delta \Gamma_\kappa[\varphi,g]}{\delta g^{\mu\nu}(x)}\right|_{\varphi_\kappa,g_\kappa}=0.
\eeq

The second equation in \eqn{eq:geom} is nothing but the set of (regularised) semiclassical Einstein equations, which encode the backreaction of the scalar field quantum fluctuations on the classical metric field $g_{\mu\nu}$. Solving the system \eqn{eq:geom} for each $\kappa$ yields the RG flow of the effective metric. One important advantage is that we can consistently study the RG flow restricted to the hypersurface of maximally symmetric field/metric configurations. This corresponds to constant field configurations $\varphi(x)=\varphi$ and, for the case of positive curvature which we consider here, to the de Sitter geometry $g_{\mu\nu}(x)=g_{\mu\nu}^{H}(x)$, characterised by the (Hubble) scale $H$. More specifically, we shall consider the expanding Poincar\'e patch of the de Sitter geometry, which, in terms of conformal time $\eta\in\mathds{R}^-$ and comoving spatial coordinates, reads $g_{\mu\nu}^{H}(x)=\eta_{\mu\nu}/(H\eta)^2$, with $\eta_{\mu\nu}$ the Minkowski metric.
Writing the effective action for constant field as $\Gamma_\kappa[\varphi, g^H]=\int_x V_\kappa(\varphi,H)$, with $V_\kappa$ the (running) effective potential, and assuming a symmetric situation with $\varphi_\kappa=0$, \Eqn{eq:geom} reduces to the semiclassical Friedmann equation $\partial_H(H^{-4}V_\kappa)|_{H_\kappa}=0$ for the (running) Hubble parameter $H_\kappa$. In the following, we are interested in the effects of superhorizon fluctuations, on scales larger than $H_{\kappa_0}^{-1}$, for a given initial value $H_{\kappa_0}$ at the initial scale $\kappa_0$. We thus choose $\kappa_0\sim H_{\kappa_0}$ and integrate the flow down to $\kappa=0$.

The physics of the massless, minimally coupled scalar field in de Sitter spacetime is well-known: Fluctuations on superhorizon scales are dramatically amplified by the gravitational field, which results in nonperturbative infrared effects. In the NPRG framework, this is manifest in the phenomenon of dimensional reduction \cite{Serreau:2013eoa,Guilleux:2015pma}: For infrared scales, the RG flow of the effective potential reduces to that of an effective zero-dimensional theory, whose solution at $\kappa=0$ is identical to the late time equilibrium state of the stochastic approach of Ref.~\cite{Starobinsky:1994bd}. Moreover, it is easy to see from the Friedmann equation that the infrared flow of the Hubble parameter is dominated by that of the effective potential, whereas derivative terms in the energy-momentum tensor give no contribution in this regime. These are dominated by ultraviolet scales \cite{Boyanovsky:2005sh} and only affect the initial value $H_{\kappa_0}$. Finally, it has been shown that the exact effective potential in the infrared limit can be obtained from the lowest order approximation in a derivative expansion of the regularized effective action, known as the local potential approximation (LPA). This has been discussed in depth in Refs.~\cite{Serreau:2013eoa,Guilleux:2015pma} to which we refer the reader for details. In the infrared regime $\kappa\ll H_\kappa$ and for field values where $\partial^2_\varphi V_\kappa\ll H^2_\kappa$, the (fully functional) flow equation for the effective potential reads [this implicitly assumes the de Sitter invariant Chernikov-Tagirov-Bunch-Davies vacuum state \cite{Chernikov:1968zm,Bunch:1978yq} for the quantum field]
\begin{equation}
    \kappa\partial_\kappa{V}_\kappa(\varphi,H) = \frac{H^4}{\Omega} \frac{\kappa^2}{\kappa^2 + \partial^2_\varphi V_\kappa(\varphi,H)},
    \label{eq:flowpot}
\end{equation}
where $\Omega=8\pi^2/3$. \Eqn{eq:flowpot} is equivalent to the standard LPA flow equation in the Euclidean space $\mathds{R}^D$ (using an appropriate regulator), with $D=0$ \cite{Delamotte:2007pf}. This effective dimensional reduction originates from the peculiar dynamics of minimally coupled massless fluctuations in a de Sitter geometry, described by $\square\hat\varphi=0$. In standard cosmological coordinates, the rapid spacetime expansion strongly washes out (comoving) spatial gradients and damps away temporal evolution on time scales larger the Hubble rate, leaving the constant zero mode as the only relevant degree of freedom.

An important consequence is that the functional integral representation of the infrared effective theory reduces to a simple integral over the single fluctuating degree of freedom left~\cite{Guilleux:2015pma}. Consider the following generating function
\begin{equation}
    e^{{\cal V}_4\mathcal{W}_\kappa(j,H)} = \int d{\hat\varphi} \,e^{-{\cal V}_4\left\{V_{\rm in}(\hat\varphi,H) + \frac{\kappa^2\hat\varphi^2}{2}-j\hat\varphi\right\}}
    \label{eqn:flowsol}
\end{equation}
where ${\cal V}_4=\Omega/H^4$, and $V_{\rm in}$ is to be specified below. It is an easy exercise to check that the modified Legendre transform $V_\kappa$, defined as $V_\kappa(\varphi,H) + \kappa^2 \varphi^2/2+\mathcal{W}_\kappa(j,H) = j\varphi$,
satisfies \Eqn{eq:flowpot} and, thus, coincides with the effective potential of our problem provided one adjusts $V_{\rm in}$ in \Eqn{eqn:flowsol} to match the initial condition at $\kappa=\kappa_0$. For large enough $\kappa_0$, we have $V_{\rm in}\approx V_{\kappa_0}$. In the following, we choose a massless, minimally coupled field described by
\beq\label{eq:Vin}
    V_{\rm in}(\hat\varphi,H)=\alpha-\beta H^2/2+\lambda\hat\varphi^4/8,
\eeq
where the $\hat\varphi$-independent terms reflect the effective Einstein-Hilbert action $M_P^2\int_x({\cal R}/2-\Lambda)$at the scale $\kappa_0$, with ${\cal R}=12H^2$ the Ricci scalar and $\Lambda$ the cosmological constant. One could add a term $H^4$ for completeness but the latter does not play any role in what follows.  The parameters $\alpha$ and $\beta$ are related to the cosmological constant $\Lambda$ and the Planck mass $M_P$ as $\alpha=\Lambda M_P^2$ and $\beta=12M_P^2$. More general cases, with nonzero mass and nonminimal gravitational coupling can be discussed along the same lines and are studied in detail in a companion paper \cite{Moreau:next}.

Using the representation \eqn{eqn:flowsol}, one easily shows that the equation $\partial_H(H^{-4}V_\kappa)|_{H_\kappa}=0$ is equivalent to 
\beq\label{eqn:minV}
   \ev{H\partial_H \qty[H^{-4}V_{\rm in}(\hat\varphi,H)]}_{\!\kappa}= 2H^{-4}\kappa^2 \ev{\hat\varphi^2}_{\!\kappa},
\eeq
where the expectation values are computed with the measure in \eqn{eqn:flowsol} evaluated at $j=0$ and $H=H_\kappa$. Using the explicit expression \eqn{eq:Vin} this rewrites as the following (regularised) semiclassical Friedmann equation
\begin{equation}
    H_\kappa^2=\frac{4}{\beta}\qty(\alpha + \frac{\kappa^2}{2}\ev{\hat \varphi^2}_{\!\kappa} + \frac\lambda8  \ev{\hat\varphi^4}_{\!\kappa}),
    \label{eqn:einsteinsc}
\end{equation}
to be compared to the classical solution $H_{\rm cl}^2=4\alpha/\beta$. Notice also that this is an implicit equation since the expectation values on the right-hand side depend on $H_\kappa$. 

We can further simplify this equation by using the identity $\int d\hat\varphi \,e^{-v(\hat\varphi)}\hat\varphi v'(\hat\varphi)=\int d\hat\varphi \,e^{-v(\hat\varphi)}$, valid for suitable functions $v$. Applied to \Eqn{eqn:flowsol}, we obtain the relation $\ev{\kappa^2\hat\varphi^2+\lambda\hat\varphi^4/2}_\kappa=H_\kappa^4/\Omega$, which allows us to eliminate the quartic term in \Eqn{eqn:einsteinsc}, and we get
\beq\label{eq:Heqfinal}
 4\alpha-\beta H_\kappa^2+H_\kappa^4/\Omega+\kappa^2\ev{\hat\varphi^2}_\kappa=0.
\eeq
Finally, before presenting explicit results, let us recall the range of validity of our approach. First, the semiclassical approximation requires that $H_\kappa^2/M_P^2\ll1$, which implies $\alpha/\beta^2\ll1$. Second, as already mentioned, the flow equation \eqn{eq:flowpot} is valid for $\kappa^2,\partial_\varphi^2V_\kappa\ll H_\kappa^2$. 

The exact flow of $H_\kappa$ is easily found numerically from Eqs.~\eqn{eqn:flowsol} and \eqn{eq:Heqfinal}. It is instructive, though, to analyse how the flow develops as $\kappa$ is decreased from the initial scale $\kappa_0$. For sufficiently large $\kappa$ the regularised theory, Eqs.~\eqn{eqn:flowsol} and \eqn{eq:Vin}, is essentially that of a nearly Gaussian field with mass $\kappa$ and perturbation theory is applicable. The two-point correlator of the Gaussian theory is $\ev{\hat\varphi^2}_{0,\kappa}=H_\kappa^4/(\Omega\kappa^2)$ and \Eqn{eq:Heqfinal} becomes, at tree level, $4\alpha - \beta H^2_{\kappa_0} + 2H^4_{\kappa_0} /\Omega= 0$. The consistency of our approach selects the solution with $H_{\kappa_0}^2/\beta\ll1$:
\begin{equation}
    H^2_{\kappa_0} \approx H^2_{\rm cl}+{2H^4_{\rm cl}}/{\beta\Omega},
    \label{eq:UV}
\end{equation}
where the second term is the first quantum correction to the classical solution $H_{\rm cl}^2=4\alpha/\beta$ at the scale $\kappa_0$. Here, we have neglected terms of relative order $H^4_{\rm cl}/(\beta\Omega)^2$.  

\begin{figure}[t]
    \centering
    \includegraphics[width=.475\textwidth]{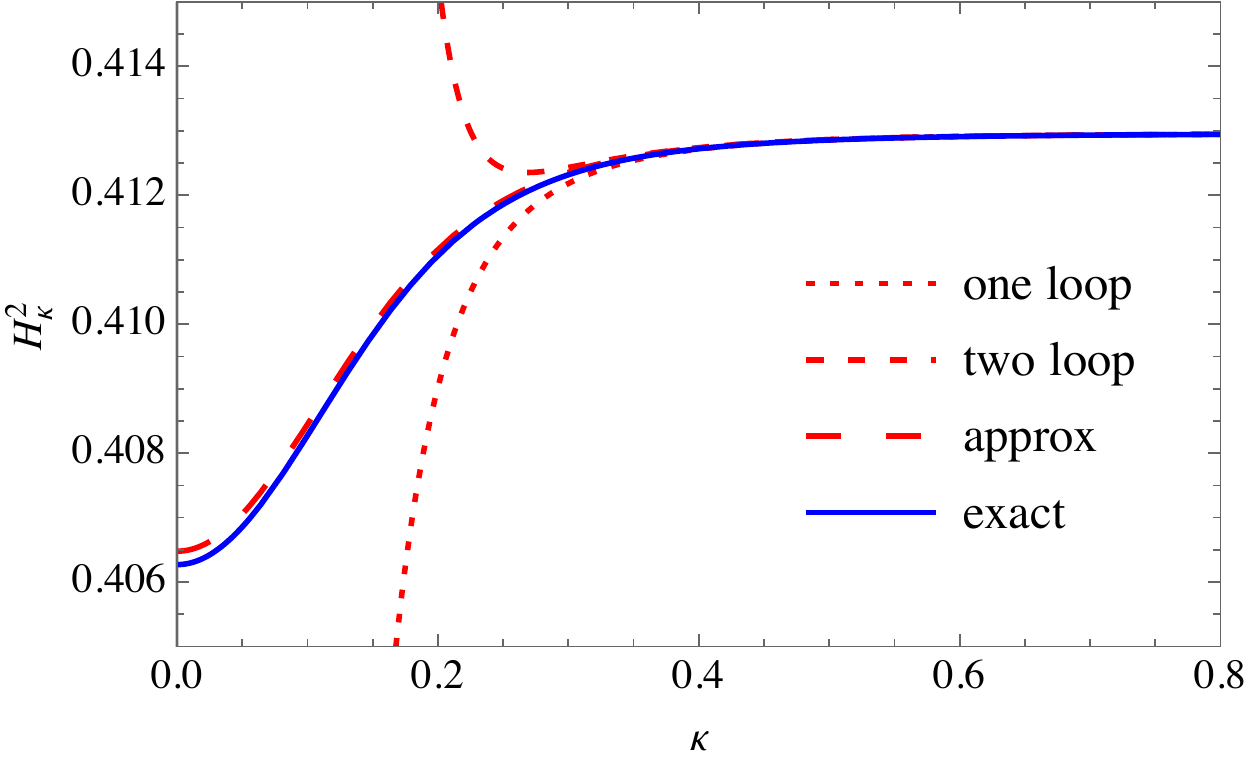}
    \caption{Flow of the spacetime curvature $H_\kappa^2$ with parameters $\alpha=0.1$, $\beta=1$, and $\lambda=0.1$. Also shown are the one- and two-loop perturbative results, which correctly describe the flow at sufficiently large $\kappa$. For lower $\kappa$, all loop orders contribute equally and perturbation theory breaks down. The flow eventually saturates  as a nonperturbative mass is dynamically generated. The long-dashed curve shows the approximate expression \eqn{eq:approx}.}
    \label{fig:flow}
\end{figure}

Deviation from the solution \eqn{eq:UV} come from perturbative corrections to the correlator $\ev{\hat\varphi^2}_\kappa$. Using Feynman diagrams, one easily checks that the actual expansion parameter is $\lambda_{{\rm eff},\kappa}=(\lambda \Omega/H_\kappa^4) \langle\hat\varphi^2\rangle_{0,\kappa}^2=\lambda H_\kappa^4/(\Omega\kappa^4)$, which is the effective coupling of the regularised zero-dimensional theory. The growth of the latter as $\kappa$ decreases is, thus, a direct consequence of the gravitational amplification of infrared Gaussian fluctuations. The first nontrivial (one-loop) correction is
\beq\label{eq:correloneloop}
 \ev{\hat\varphi^2}_\kappa=\ev{\hat\varphi^2}_{0,\kappa}\qty[1-3\lambda_{{\rm eff},\kappa}/2+{\cal O}(\lambda_{{\rm eff},\kappa}^2)],
\eeq
from which we obtain, defining $\tilde\beta=\beta\Omega/H_{\kappa_0}^2$,
\beq
 H_\kappa^2/H_{\kappa_0}^2=1-3\lambda_{{\rm eff},\kappa}/(2\tilde\beta)+{\cal O}(\lambda_{{\rm eff},\kappa}^2,\tilde\beta^{-2}).
\eeq
We see that the one-loop contribution leads to a decay of the spacetime curvature as superhorizon fluctuations are progressively integrated out, as shown in \Fig{fig:flow}.

However, when the one-loop correction becomes significant, higher orders also become important and, in fact, the perturbative expansion breaks down, as also illustrated in  \Fig{fig:flow}. In this nonperturbative regime, large infrared fluctuations trigger the dynamical generation of an effective mass, which screens the growth of quantum fluctuations on deep superhorizon scales \cite{Starobinsky:1994bd,Guilleux:2015pma}. The correlator $\ev{\hat\varphi^2}_{\kappa=0}$ is thus finite and can be trivially evaluated from \Eqn{eqn:flowsol} as
\beq
 \ev{\hat\varphi^2}_{\kappa=0}=\frac{\Gamma(3/4)}{\Gamma(1/4)}\sqrt{\frac{8H_\kappa^4}{\lambda\Omega}}.
\eeq
It follows from \Eqn{eq:Heqfinal} that $H_\kappa^2$ saturates to a finite value at $\kappa=0$, given by $4\alpha - \beta H_{\kappa=0}^2 +H^4_{\kappa=0}/\Omega = 0$, that is,
\beq
  H^2_{\kappa=0}\approx H^2_{\rm cl}+{H^4_{\rm cl}}/{\beta\Omega}.
    \label{eq:IR}
\eeq
The relative change in $H_\kappa^2$ is thus given by \mbox{$ H_{\rm cl}^2/(\beta\Omega)\ll1$.} 

The complete flow is obtained by solving \Eqn{eq:Heqfinal} with the exact expression of the correlator [$K_\nu(x)$ is the modified Bessel function of the second kind]
\beq\label{eq:correl}
 \frac{\ev{\hat\varphi^2}_\kappa}{\ev{\hat\varphi^2}_{0,\kappa}}=\lambda_{{\rm eff},\kappa}^{-1}\qty[\frac{K_{3\over4}(\lambda_{{\rm eff},\kappa}^{-1}/4)}{K_{1\over4}(\lambda_{{\rm eff},\kappa}^{-1}/4)}-1]\equiv {\cal C}_\kappa(H_\kappa),
\eeq
which implicitly depends on $H_\kappa$ through $\lambda_{{\rm eff},\kappa}$. An approximate explicit solution of \Eqn{eq:Heqfinal} can be obtained by expanding around $H_{\kappa_0}^2$ in inverse powers of $\tilde\beta$:
\beq\label{eq:approx}
 {H_\kappa^2/H_{\kappa_0}^2}=1-\tilde\beta^{-1}\qty[1-{\cal C}_\kappa(H_{\kappa_0})]+{\cal O}(\tilde \beta^{-2}),
\eeq
where function ${\cal C}_\kappa$ is defined in \Eqn{eq:correl}. This is shown in \Fig{fig:flow} together with the exact solution and the breakdown of the perturbative expansion.

In conclusion, we have investigated the backreaction of a light quantum scalar field on a de Sitter geometry by means of recently developed NPRG techniques. We find a nontrivial renormalisation of the spacetime curvature as superhorizon fluctutations are progressively integrated out. Perturbative loop corrections grow unbounded as a result of the gravitational amplification of such fluctuations. This signals the breakdown of perturbation theory rather than an instability. Nonperturbative effects come into play with, in particular, the dynamical generation of a mass, which screens the growth of superhorizon fluctuations and freezes the RG flow of the effective spacetime curvature. Overall, the infrared renormalisation of the latter is controlled by the gravitational coupling $H_{\kappa_0}^2/(\beta\Omega)\approx\Lambda/(96\pi^2M_P^2)$, small by assumption in the present semiclassical treatment.

We believe the present work brings an interesting light on the general issue of de Sitter spacetime stability against quantum fluctuations. It is worth emphasising, though, that, here, we have discussed the specific case of superhorizon fluctuations of a quantum scalar field in a de Sitter invariant quantum state. Various other possible sources/directions of instability have been discussed in the literature such as, for instance, the question of stability of a global de Sitter geometry (as opposed to the expanding Poincar\'e patch studied here) \cite{Polyakov:2007mm,Akhmedov:2012dn}, non-de Sitter-symmetric quantum states for the scalar field \cite{Boyanovsky:2011xn,Anderson:2013ila,Akhmedov:2013xka,Anderson:2017hts,Akhmedov:2017ooy}, or the role of graviton fluctuations \cite{Tsamis:1992sx,Miao:2017vly}. It remains to be investigated whether the NPRG techniques proposed here can be useful in such cases.

\end{document}